\documentclass[10pt,   showpacs, amsmath,amsfonts, amssymb,eqsecnum,  twocolumn, prd]{revtex4}%twocolumn, draft,
\usepackage{epsfig,eucal}

 \usepackage{amsfonts,amssymb}
\usepackage{amsmath}
\usepackage{youngtab}
\usepackage{graphicx}

\def\a{\alpha}

\def\ep{\varepsilon}

\def\a{\alpha}

\begin{document}
 %\maketitle
%----------------------------
  \title{Electromagnetic media with Higgs-type spontaneously  broken  transparency}
\author{Yakov Itin\\Institute of Mathematics, The Hebrew University of
 Jerusalem \\ and Jerusalem College of Technology, Jerusalem,
  Israel. \\ email: {\tt itin@math.huji.ac.il}}

%\affiliation{Institute of Mathematics, The Hebrew University of
%  Jerusalem \\ and Jerusalem College of Technology, Jerusalem,
%  Israel. \\ email: {\tt itin@math.huji.ac.il}}

 %---------------------------headings------------
\pagestyle{myheadings}
\markboth{Yakov Itin} {Yakov Itin \qquad\qquad\qquad\qquad\qquad\qquad\qquad\qquad\qquad\qquad
{Electromagnetic media with spontaneous breaking transparency}}
%Optic tensors and skewon. 

 %---------------------------headings------------
%\pagestyle{myheadings}
%\markboth{Yakov Itin} {Yakov Itin \qquad
%On light propagation  in skewon media}

%\markboth
%\begin{document}
%%%%%%%%\maketitle
 \begin{abstract}
In the framework of  standard electrodynamics with  linear local response, we construct a model that provides spontaneously broken   transparency. The functional dependence of the medium parameter turns out to be of the Higgs type. 
\end{abstract}
%\keywords{Electrodynamics; Relativity; Constitutive law; 
%Anisotropic media}
\date{\today}
\maketitle
%\tableofcontents

%--------------------------------------------
\section{Introduction }
%--------------------------------------------
Physical models often involve phenomenological parameters or auxiliary fields characterizing the background spacetime or the background media. In most cases, dynamics of the model depend smoothly (continuously and differentiably) on the values of the background parameter. A non-smooth functional dependence is a rather rare phenomenon, but if it exists, it usually represents a keystone issue of the model. The examples of such non-smooth behavior are well known in solid state physics as phase transitions at  critical points. Another similar issue is the scalar  Higgs  model of  spontaneous symmetry breaking. 

In this paper, we present a simple phenomenological model of  an electromagnetic medium that allows wave propagation only for a sufficiently big value of the medium parameter. For zero values of the parameter, our medium is the ordinary SR (or even GR) vacuum with the standard dispersion relation $\omega^2=k^2$. However  even  infinitesimally small variations of the parameter modify the dispersion relation in such a way that it does not have real solutions, i.e.,  the medium becomes to be completely opaque. For  higher values of the parameter, the dispersion relation is modified once more and once again it has real solutions. It is well known that the dispersion relation can be treated as an effective metric in the phase space. In our model, the vacuum Lorentz metric is spontaneously transformed  into the Euclidean one and returns to be Lorentzian for a sufficiently big value of the parameter. 

\section{Skewon modified electrodynamics}
We consider the standard electromagnetic system of two antisymmetric fields $F_{ij}$ and $H^{ij}$ that obey the vacuum Maxwell system
\begin{equation}\label{Max}
F_{[ij,k]}=0\,,\qquad H^{ij}{}_{,j}=0\,.
\end{equation}
The fields are assumed to be related by the  local linear  constitutive relation, \cite{Post},\cite{birkbook},
\begin{equation}\label{Cons}
H^{ij}=\frac 12 \chi^{ijkl}F_{kl}\,.
\end{equation}
Due to this definition, the constitutive tensor obeys the symmetries 
\begin{equation}\label{Sym}
\chi^{ijkl}=-\chi^{jikl}=-\chi^{ijlk}\,.
\end{equation}
The electromagnetic model (\ref{Max}) with the local linear response  (\ref{Cons})  is intensively studied recently, see \cite{Obukhov:2000nw}, \cite{Obukhov:2002xa}, \cite{Lammerzahl:2004ww}, and especially in \cite{birkbook}. 

By using the  Young diagram technique,  a fourth rank tensor with the symmetries (\ref{Sym}) is uniquely irreducible decomposed into the sum of three independent pieces.
  \begin{equation}\label{Decomp}
\chi^{ijkl}={}^{\tt (1)}\chi^{ijkl}+{}^{\tt (2)}\chi^{ijkl}+{}^{\tt(3)}\chi^{ijkl}\,.
\end{equation}
The first term here is the principal part. In the simplest pure Maxwell  case it is expressed by the metric tensor of GR
\begin{equation}\label{Princ-Part}  
{}^{(1)}\chi^{ijkl}=\sqrt{|g|}\left(g^{ik}g^{jl}-g^{il}g^{jk}\right)\,.
\end{equation}
In the flat Minkowski spacetime with the metric $\eta^{ij}={\rm diag}(1,-1,-1,-1)$, it reads
\begin{equation}\label{Princ-Part-M}  
{}^{(1)}\chi^{ijkl}=\eta^{ik}\eta^{jl}-
\eta^{il}\eta^{jk}\,.
\end{equation}
In quantum field description, this term is related to the  photon. 

The third term in (\ref{Decomp}) is completely skew symmetric. Consequently, it can be written as 
\begin{equation}\label{Axion-Part}  
{}^{\tt(3)}\chi^{ijkl}=\a\ep^{ijkl}\,.
\end{equation}
The pseudo-scalar $\a$ represents the axion copartner of the photon. It influences  the wave propagation such that birefringence occurs  \cite{Ni:1977zz}, \cite{Carroll:1989vb}. In fact, this effect is absent in the geometric optics description and corresponds to the higher order approximation, \cite{Itin:2004za}, \cite{Itin:2007wz}, \cite{Itin:2007cv}. 

We turn now to the second part of (\ref{Decomp}), that is expressed as 
\begin{equation}\label{Skewon-Part}  
{}^{\tt(2)}\chi^{ijkl}=\frac 12\left(\chi^{ijkl}-\chi^{klij}\right)\,.
\end{equation}
This tensor has 15 independent components, so it may be represented by a traceless matrix \cite{birkbook}, \cite{Obukhov:2004zz}. 
This matrix reads
\begin{equation}\label{Skewon-Matr}  
S_i{}^j=\frac 14 \ep_{iklm}{}^{\tt(2)}\chi^{klmj}\,.
\end{equation}
The traceless condition $S_k{}^k=S_{ij}g^{ij}=0$ follows straightforwardly from (\ref{Skewon-Matr}). 

In order to describe the influence of the skewon on the wave propagation, it is convenient to introduce a covector 
\begin{equation}\label{Skewon-Cov}  
Y_i=S_i{}^jq_j\,.
\end{equation}
Consider a medium described by a vacuum principal part (\ref{Princ-Part-M}) and a generic skewon. 
The  dispersion relation for such a medium takes the form,  \cite{Itin:2013ica}, \cite{Itin}, 
\begin{equation}\label{Disp}  
q^4=q^2Y^2-<\!q,Y\!>^2\,.
\end{equation}
Here the scalar product $<\!q,Y\!>$ and the squares of the covectors $q^2$ and $Y^2$ are calculated by the use of the metric tensor. 

It can be easily checked that Eq.(\ref{Disp}) is invariant under the gauge transformation 
 \begin{equation}\label{Gauge}  
Y\to Y+Cq\,,
\end{equation}
with an arbitrary real parameter $C$. This parameter can even be an arbitrary function of $q$ and of the medium parameters $C=C(q,S)$. With this gauge freedom, we can apply the Lorenz-type gauge condition $<\!q,Y\!>=0$ and obtain the dispersion relation in an even more simple form 
\begin{equation}\label{Disp-Short}  
q^4=q^2Y^2\,.
\end{equation}
This expression yields a characteristic fact \cite{Itin:2013ica}: The solutions $q_i$ of the dispersion relation, if they exist, are non-timelike, that is, spacelike or null, 
\begin{equation}\label{Non-Time}  
q^2\le 0\,.
\end{equation} 

We will proceed now with the  form (\ref{Disp}) and with the skewon covector expressed as in (\ref{Skewon-Cov}). 
We can rewrite the dispersion relation as 
\begin{equation}\label{Sol}  
q^2=\frac 12 \left(Y^2\pm\sqrt{Y^4-4<\!q,Y\!>^2}\right)\,.
\end{equation} 
Consequently, the real solutions  exist only if 
\begin{equation}\label{Ineq}  
0\le Y^4-4<\!q,Y\!\!>^2\,.
\end{equation} 
Our crucial observation that the first term here is quartic in the skewon parameters $S_{ij}$ while the second term is only quadratic. Under these circumstances, the first term can be small  for  for  sufficiently small skewon parameters  and the inequality (\ref{Ineq}) breaks down. For higher values, the first term  becomes to be essential and  the  inequality is  reinstated. 

\section{A model}
We now present a model where this possibility is  realized, indeed. Consider a symmetric traceless matrix with two nonzero entries
\begin{equation}\label{Matrix}  
S_{00}=S_{11}=\sigma\,.
\end{equation}
We denote  the components of the wave covector as $q_i=(\omega,{\rm k}_1,{\rm k}_2,{\rm k}_3)$. 
The skewon covector has two nonzero components
 \begin{equation}\label{Matrix}  
Y_0=\sigma \omega\,,\qquad Y_1=-\sigma {\rm k}_1\,.
\end{equation}
Consequently,
\begin{equation}\label{Calc}  
Y^2=\sigma^2 \left(\omega^2-{\rm k}_1^2\right) \,,\quad <\!q,Y\!\!>=\sigma\left(\omega^2+{\rm k}_1^2\right) \,.
\end{equation}
Hence the inequality (\ref{Ineq}) takes the form
\begin{equation}\label{Ineq1}  
\sigma^4 \left(\omega^2-{\rm k}_1^2\right)^2-4\sigma^2\left(w^2+{\rm k}_1^2\right)^2\ge 0\,.
\end{equation} 
Observe that for every choice of the wave covector this expression is of the form $f(\sigma)=A\sigma^4-B\sigma^2$ with positive coefficients $A,B$. Quite surprisingly, this functional expression repeats the well known curve of the  Higgs  potential. 
\begin{figure}[h!]
{
\includegraphics[width=6.5cm]{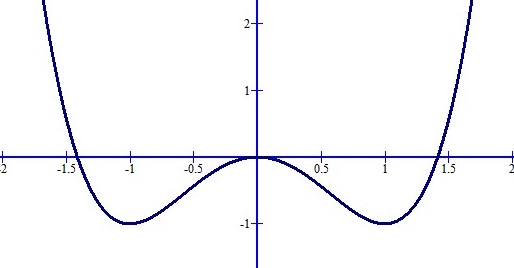}}
\caption[]{Function $f(\sigma)=A\sigma^4-B\sigma^2$.}
%}
\end{figure}

The dispersion relation as it is given in  Eq.(\ref{Disp}) reads
\begin{equation}\label{Disp1}  
q^4-q^2\sigma^2\left(\omega^2-{\rm k}_1^2 \right) +\sigma^2\left(\omega^2+{\rm k}_1^2 \right)^2 =0\,.
\end{equation} 
We rewrite it as 
\begin{eqnarray}\label{Disp2}  
&&\left(q^2- \frac{\sigma^2}2\left(\omega^2-{\rm k}_1^2\right) \right)^2 +4\sigma^2\omega^2k_1^2+\nonumber\\&&\qquad\qquad\qquad
\frac{\sigma^2}4\left(4-\sigma^2\right)\left(\omega^2-{\rm k}_1^2 \right)^2=0\,.
\end{eqnarray}

Consequently:
\begin{itemize}
\item[(i)] For $\sigma=0$, we return  to the unmodified light cone $q^2=0$.

\item[(ii)]  For $0<|\sigma|\le 2$, except for the trivial solution $q_i=0$, there are no real solutions of Eq(\ref{Disp2}) at all. 

\item[(iii)] For $|\sigma|> 2$, there  are two real solutions: 
\end{itemize}
 \begin{equation}\label{sym-15}
q^2=\frac{\sigma^2}2\left(\omega^2-{\rm k}_1^2 \right) \pm \frac{\sigma}2\sqrt{\left(\sigma^2-4\right)\left(\omega^2-{\rm k}_1^2\right)^2-16\omega^2k_1^2 }%\,.
\end{equation}
For the numerical images of these algebraic cones,  see Fig. 3 and Fig. 4. 
  \begin{figure}[h!]
{
\includegraphics[width=4.5cm]{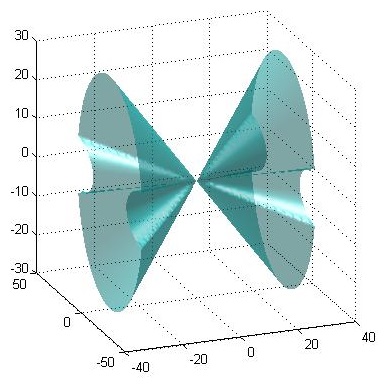}%}%\qquad
%{
\includegraphics[width=4.5cm]{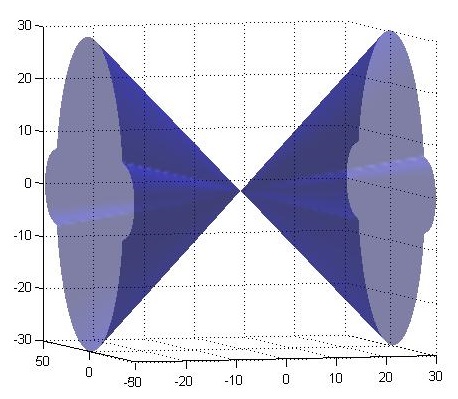}}
\caption[]{Algebraic cones of the rank 2 symmetric skewon (\ref{sym-15}) with $"-"$  and $"+"$ signs respectively.  The parameter $\sigma=\sqrt{5}$. $\omega$ is directed as $z$-axis, $k_1,k_2$ are directed as $x$ and $y$ axes respectively. $k_3=0$.  }
%}
\end{figure}
\begin{figure}[h!]
{
\includegraphics[width=4.5cm]{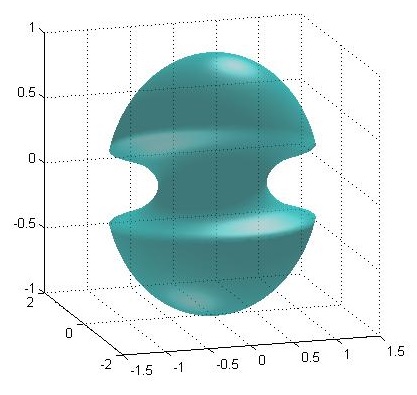}%}%\qquad
%{
\includegraphics[width=4.5cm]{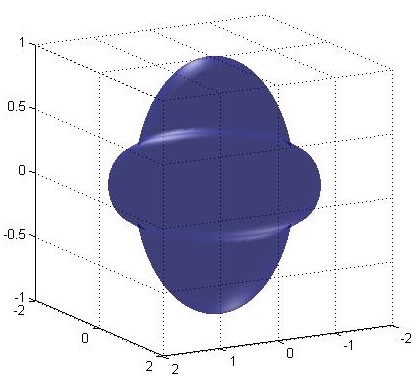}}
\caption[]{Wave front for the rank 2 symmetric skewon (\ref{sym-15}) with $"-"$  and $"+"$ signs respectively.  The parameter $\sigma=\sqrt{5}$. $\omega$ is directed as $z$-axis, $k_2,k_3$ are directed as $x$ and $y$ axes, respectively. $k_3=1$.  }
%}
\end{figure}
 % \begin{figure}[h!]
%\parbox [t ]{0.5\textwidth }{
%\includegraphics[width=8.1cm]{rank2_1y}}
%\caption[]{Skewon wit two light cones. The parameter $A=4$.}
%}
%\end{figure}
In both cones, the skewon interchanges the time axis with the spatial $x$-axis. 
These 3-dimensional  cones are tangential to one another  when the discriminant in  Eq. (\ref{sym-15}) is zero. It gives a 2-dimensional cone 
 \begin{equation}\label{sym-15x}
{\rm k}_1^2=\frac{\sigma+2}{\sigma-2}\omega^2\,,\qquad 
{\rm k}_2^2+{\rm k}_3^2=2\frac{\sigma^2-2}{\sigma-2}\omega^2\,.
\end{equation}

 The expressions in Eq.(\ref{sym-15}) can be treated as Finlsler metric elements. Due to the fact that they have non-compact 2D-sections as in Fig. 2 and compact 2D-sections as in Fig. 3, the corresponding Finsler metric tensors are of the Lorentz signature type. 

We construct a model of the electromagnetic vacuum with   a skewon field that has the following features:
\begin{itemize}
\item[(1)] 
There is a gap for values of the parameters near zero, where the wave propagation is forbidden. 
\item[(2)] Birefringence of the light propagation. 
\item[(3)]  Full interchange between time and spatial direction. 
\item[(4)] A continuous 2-dimensional variety of  optic axes instead of  distinct optic axes appearing in anisotropic optics. 
\item[(5)] The light cones are non-convex.
\end{itemize} 
\section{Conclusion}
Electromagnetic media with   an additional skewon field provide a rich class of models of wave propagation with rather unusual features, see \cite{Obukhov:2004zz}, \cite{Itin}. Recently the observational restrictions on such models were discussed in \cite{Ni:2013uwa} and \cite{Alighieri:2014yoa}. 
In this paper, we show that Higgs-type potential can appear in a simple electromagnetic model  by a minimal modification of the vacuum constitutive relation.

%-----------------------------
\section*{Acknowledgments}
My acknowledgments to F. Hehl (Cologne/Columbia, MO), Yu. Obukhov (Cologne/Moscow), V. Perlick (ZARM, Bremen), C. L\"ammerzahl (ZARM, Bremen), and Y. Friedman (JCT, Jerusalem) 
for valuable discussions.  I acknowledge  the GIF grant No. 1078-107.14/2009 for financial support.

\end{document}